# Unidirectional dissipative soliton operation in an all-normal-dispersion bidirectional Yb-doped fiber laser without an isolator

Daojing Li,[1] Deyuan Shen,[1] Lei Li,[2] Hao Chen,[2,*] Dingyuan Tang,[2] Luming Zhao[2]

[1] *Department of Optical Science and Engineering, Fudan University, Shanghai 200433, China*
[2] *Jiangsu Key Laboratory of Advanced Laser Materials and Devices,*
*School of Physics and Electronic Engineering, Jiangsu Normal University, Xuzhou, Jiangsu, 221116, China*
*\*Corresponding author: chenhao@jsnu.edu.cn*

We demonstrate self-started unidirectional dissipative soliton operation and noise-like pulse operation in an all-normal-dispersion bidirectional Yb-doped fiber laser mode-locked by nonlinear polarization rotation. The laser works unidirectional once mode locking was achieved due to the cavity directional nonlinearity asymmetry along with the nonlinear polarization rotation mode locking mechanism.

## 1. Introduction

Passively mode-locked fiber lasers have been extensively investigated as an excellent ultrafast pulse source because of their compactness, alignment-free operation and excellent pulse stability [1-9]. Dissipative solitons (DSs) were recently observed in all-normal-dispersion fiber lasers [1-3], which become a good alternative to solid-state lasers as they favor larger pulse energy compared to the solitons formed in the anomalous-dispersion regime and the dispersion-managed regime [4,5]. Generally, a saturable absorber (SA) is required to achieve passively mode locking. Either physical or artificial SA can be employed in the cavity. The former includes semiconductor saturable absorber mirrors (SESAMs), carbon nanotubes (CNTs) and graphene etc., while the latter includes nonlinear polarization rotation (NPR), nonlinear optical loop mirror (NOLM), and their variants.

It has been shown that a unidirectional ring cavity configuration could reduce the spurious cavity reflections and decrease the mode-locking self-starting threshold. Tamura *et al.* experimentally demonstrated that the mode locking threshold power could be significantly decreased in a unidirectional ring cavity than a linear cavity [6]. In the early days of fiber laser development, due to the lack of high power pump sources, unidirectional ring cavity configuration was mostly adopt for mode-locked fiber lasers. Conventionally, an isolator was used to ensure the unidirectional operation. With a high power pump, Zhao *et al.* reported self-started mode-locking and soliton operation in a bidirectional fiber ring laser mode locked by the NPR technique [7]. It was shown that due to the cavity nonlinearity-induced directional symmetry breaking the laser always operated unidirectional after mode-locking. Kieu *et al.* reported an all-fiber bidirectional ring laser mode locked by carbon nanotubes/polymer composite [8]. The laser generated two stable conventional soliton trains in opposite directions. Yao *et al.* generated bidirectional stretched pulses in a nanotube-mode-locked fiber laser [9]. The aforementioned bidirectional fiber lasers were all performed in the net anomalous region. It is well known that for conventional solitons and stretched solitons in anomalous dispersion fiber laser, the pulse shaping is mainly due to the natural balance between the anomalous dispersion and the fiber nonlinearity, whereas DSs in the normal dispersion region are the result of the combined effects among the cavity dispersion, fiber nonlinearity, gain and loss, and spectral filtering. And the dissipative effects play a crucial part on the soliton shaping. Consequently, comparing to conventional soliton and stretched soliton, DS has a relatively wider pulse duration and lower peak power. So it is desired to study the DS operation in a bidirectional fiber laser, especially when the laser is mode-locked by the NPR technique, since the NPR mode-locking has a fast recovery time and needs high peak power to accumulate enough nonlinear phase shift to start mode-locking.

In this paper, we experimentally study DS operation in an all-normal-dispersion bidirectional Yb-doped fiber (YDF) laser mode locked by the NPR. We show that self-started DS operation can still be achieved in the bidirectional fiber laser. Once mode locked, the laser operates in the clockwise (CW) direction. We attribute those features to the cavity directional nonlinearity asymmetry along with the NPR mode-locking technique.

## 2. Experimental Setup

The fiber laser is shown in Fig. 1. It had a ring cavity of about 11.7 m, including the free space distance. No isolator was introduced hence it was a bidirectional cavity. Apart from the 32 cm YDF (YB406, CorActive), all the other fibers used were standard single mode fiber (SMF). The laser was mode-locked by the NPR technique. To this end, one polarization beam splitter (PBS) together with two quarter-wave plates and one half-wave plate were used and mounted on a fiber bench. The insertion loss of the fiber bench was 1.1 dB. The laser was pumped by a 976 nm pump laser through a wavelength division multiplexer (WDM). A 10% fiber coupler was used to output the laser.

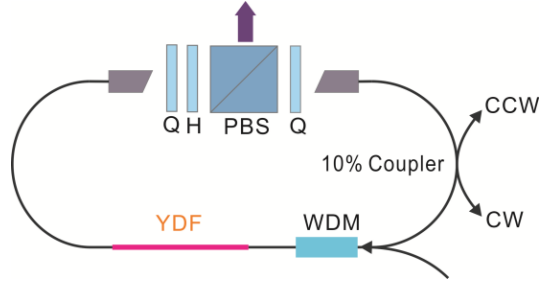

Fig. 1. Schematic of the fiber laser. Q: quarter-wave plate; H: have-wave plate; PBS: polarization beam splitter; CW: clockwise; CCW: counterclockwise. YDF: Yb-doped fiber; WDM: wavelength division multiplexer.

## 3. Experimental Results and Discussion

Self-starting mode locking could be achieved by simply rotating the wave plates, when the pump power was slowly increased to 490 mW. In continuous wave operation the laser emitted bidirectionally. However, the laser operated in the CW direction after mode locking. The detail features of the CW output are shown in Fig. 2. The pulse spectrum centered in 1062 nm exhibited characteristic steep edges, with a full width half maximum (FWHM) of 2.18 nm.  The small hump located in 1110nm was the weak Raman conversion of the DS. Figure 2(b) is the oscilloscope trace of the CW output train. The repetition rate was 17.82 MHz. The power of the CW output was 29.5 mW, corresponding to 1.66 nJ. The radio frequency (RF) spectrum is shown in Fig. 2(c), with a high contrast of near 80 dB, indicating low-amplitude fluctuations, when resolution bandwidth (RBW) of the RF spectrum analyzer is set at 10 Hz. The inset shows the RF spectrum with 100 MHz bandwidth and the RBW is set at 1 kHz. The autocorrelation trace of the CW output exhibited a triangle shape. The measured pulse duration was 36.5 ps (Gaussian shape was assumed). The pulse was highly chirped and could be compressed outside the cavity by a pair of gratings down to 1.14 ps (a time-bandwidth product of 0.66).

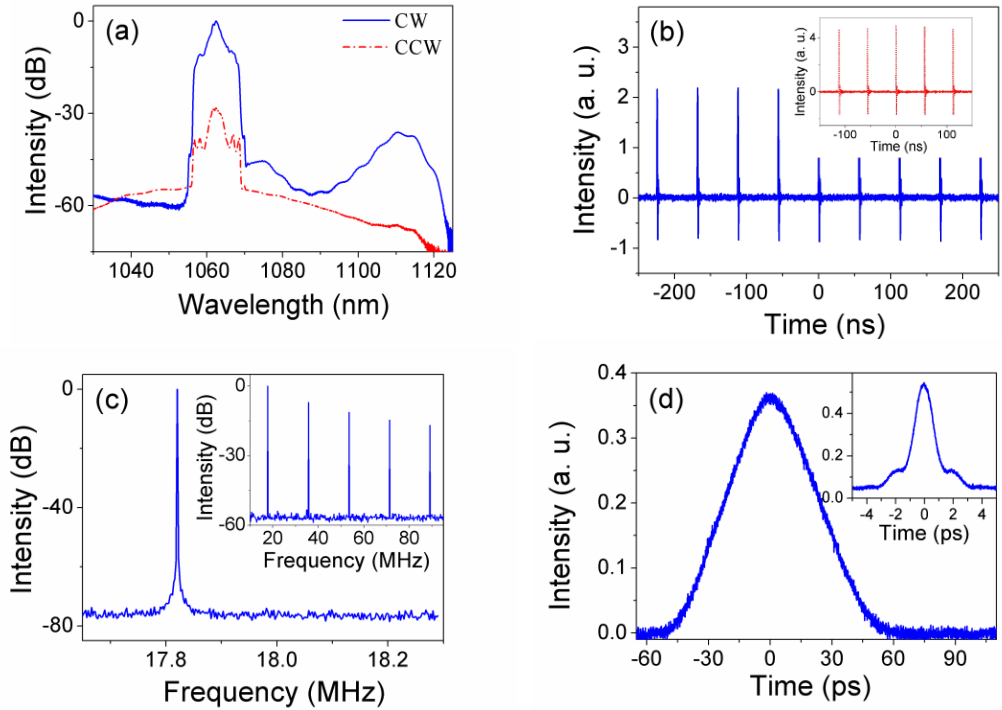

Fig. 2.  (a) Optical spectra, (b) Pulse sequences, (c) Radio frequency, (d) Chirped and dechirped (inset) pulse autocorrelation traces of the clockwise output (solid blue line) and the counterclockwise output (dashed red line).

Experimentally, apart from the DS operation in the CW direction, a weak signal was also detected from the counterclockwise (CCW) output. The CCW output power was 30 μW, around 30 dB weaker than that of the CW output. Its spectral profile was plotted in Fig. 2(a), and oscilloscope trace in the inset of Fig. 2(b). As one can see, the CCW output also worked in pulsed regime. Careful examination on the spectral profile measured reveals that it had almost the same spectral profile pattern with the CW output, sitting upon spontaneous emitting spectrum. Thus we believe that the CCW pulse was only a weak reflection of the CW pulse in the cavity.

For further research on the nature of the CCW output, we carefully rotated one wave plate and a different mode-locking operation state was obtained. Fig. 3 shows spectra of the CW output, CCW output and CCW amplified spontaneous emission (ASE) spectrum under low pump power. Again, spectra of two directions shown resemblance. In this case, stronger Raman conversion of the DS was observed in both CW and CCW output. The CCW output was pulsed as well. However, the CCW output power here (~28 μW) was also near 30 dB weaker than the CW output power (~26 mW). For such low optical power, Raman conversion cannot be intrigued. Therefore, we believe that the CCW output was not real mode-locked but reflection of the CW laser. Fig. 3(d) shows the triangle shape autocorrelation trace of the PBS output, implying a pulse duration of 30.4 ps. The pulse can be dechirped to 813 fs (a time-bandwidth product of 0.60).

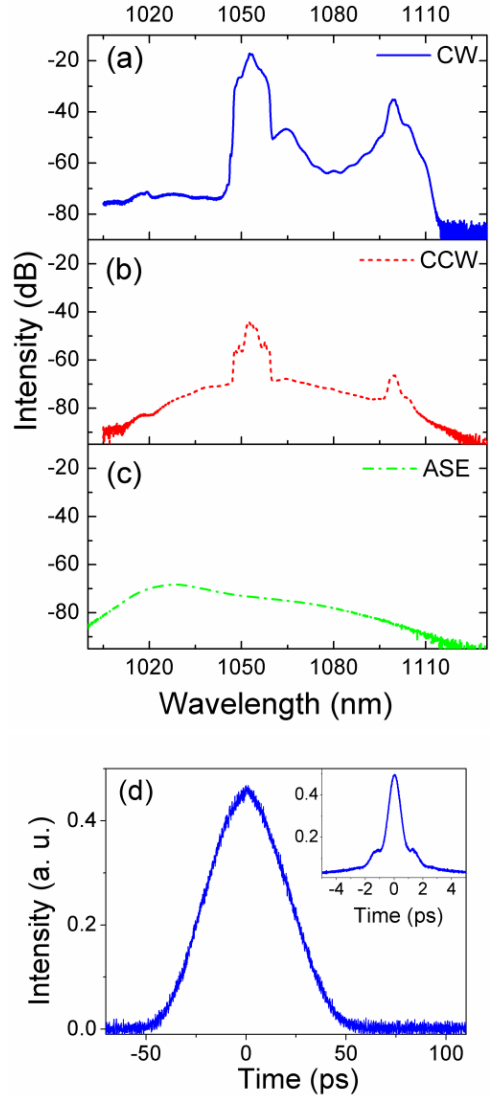

Fig. 3. Optical spectra of the (a). Clockwise output (solid blue line), (b). Counterclockwise output (dashed red line), and (c). Counterclockwise ASE spectrum (dash dotted green line). (d). Autocorrelation trace of chirped and dechirped (inset) clockwise output.

This unidirectional operation could be explained as a consequence of different nonlinearity experienced by pulses along the different directions. Let us estimate the different nonlinear phase shift accumulated along two directions. The nonlinear phase shift accumulated is given by $\Phi_{NL} = \gamma PL$, where $\gamma$ is the fiber nonlinear coefficient, $P$ is the optical power and $L$ is the distance. Carefully analyzing the laser cavity configuration, the amplified pulse in the CW direction, travels along a 6.25 m SMF, then arrives at the PBS, where the saturable absorption takes place; while the pulse in the count-clockwise (CCW) direction after being amplified in the YDF first propagates along a 3 m SMF then experiences the cavity output loss, and a 2 m SMF before reaching the PBS. Assuming laser has same pulse power after amplified at two directions and neglecting the fiber loss, the CW nonlinear phase shift is $\Phi_{CWNL} = 6.25\gamma P$, while the CCW nonlinear phase shift is $\Phi_{CCWNL} = 3\gamma P + 2\gamma*0.9P = 4.8\gamma P$. The CW laser accumulates about 30% higher nonlinear phase shift than the CCW laser. Although the laser backward amplified in the CCW direction could has higher power, we believe it would not be able to reverse the nonlinearity asymmetry. Since the NPR was adopted as mode-locking technique, light in the CW direction with larger nonlinear phase shift has lower mode-locking threshold and thus is easier to start mode locking while increasing the pump power. Therefore, this cavity directional nonlinearity asymmetry leads to the mode-locking priority.

We notice that in the formerly reported mode locking bidirectional fiber laser cavities [8,9], the lasers generated bidirectional soliton pulses. However in this work, the laser operation keeps unidirectional regardless of the increasing pump. At the maximum pump power of 1150 mW, the DS collapsed into noise-like pulse (NLP) due to the Raman scattering, which will be discussed elsewhere. For noise-like pulse operation, the laser still worked unidirectional and pulsed. Both the CW and CCW output spectra were displayed in Fig. 4(a). The autocorrelation trace of the CW output was shown in Fig. 4(b). The grooves at the right side of autocorrelation trace were due to our autocorrelator itself. The trace exhibited a femtosecond coherent spike on the top of a broad pedestal, a typical autocorrelation profile of NLPs. The CCW output power here was also around 30 dB weaker than its counterpart. And again two spectral profile shown resemblance, further proving that the CCW pulse was a weak image of the CW pulse.

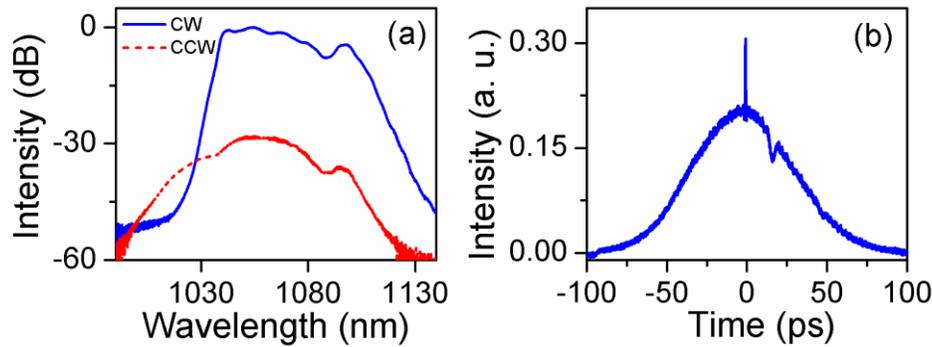

Fig. 4 (a). Optical Spectra and (b). Autocorrelation trace of noise like pulses (solid blue line, clockwise output; dashed red line, counterclockwise output)

The different operation states between this cavity and those in [8,9] is mainly due to the different mode-locking technique used. In [8,9], the lasers were mode-locked by nanotubes. The saturable absorption of nanotubes is related to light power but not sensitive to its direction. Once light in the direction with priority saturated the nanotubes and mode locked first, light in the other direction could also be saturable absorbed. Small spontaneous emission was able to start mode locking. However, the situation of the NPR adopted in this cavity is quite different. The saturable absorption of the NPR mode locking is based on accumulated nonlinear phase shift during propagation along the fiber. As light in the CW direction first mode locked, cavity energy was concentrated on pulses in CW direction. Small spontaneous emission in the CCW direction could not accumulate enough nonlinear phase shift to start mode-locking. Moreover, NPR mode-locking also depends on the appropriate intra-cavity polarization control. When the wave plates are set to achieve mode-locking for the CW laser, it not necessarily has saturable absorption for the CCW laser. Consequently, despite of pump increase, the laser operated unidirectional in the CW direction.

Any optical isolators have unavoidably insertion loss. This laser setup without isolator is promising for energy scaling of mode-locked lasers. The DS in this laser is generated with the intrinsic birefringence filter of the NPR mode-locking, which has been reported by Zhao *et al.* [10]. These two laser setups are similar, except for the absence of the isolator. In [10], the laser mode-locking threshold was about 100 mW, and the 10% output power of the generated DS was 0.72 mW at the pump power of 148 mW, which corresponds to pulse energy of 31.7 pJ. For this laser setup, unidirectional operation in the CW direction was always obtained through carefully tuning the intra-cavity polarization controller while gradually increasing the pump power, although the mode-locking threshold was increased up to 490 mW. However, DSs with 29.5 mW average power were generated under 490 mW pump, resulting much higher optical-to-optical efficiency (~6.0%) than that in [10] (~0.49%).

We notice that the laser could theoretically be able to operate in the counterclockwise direction under sufficiently high pump power. However in practice the CW mode-locking operation was always first obtained when the pump power was slowly increased. After mode-locked, wave plates were fixed and the laser keeps unidirectional regardless of pump power increase. In this way, operation direction of the laser was actually predefined through the cavity nonlinearity asymmetry design. The long-term stability of unidirectional operation was tested. We carried out repeating scans of DS spectrum in Fig. 3 with 10-min interval, and recorded central wavelength and peak level within an hour. The results is shown in Fig. 5. No significant wavelength shift was observed and the intensity fluctuation was less than 0.1 dB, indicating good mode-locking operation stability at room temperature. As the laser was outputted through a piece of SMF, the laser beam is near-diffraction-limited, the same as other SMF lasers.

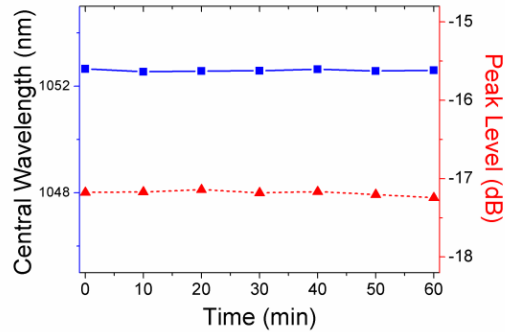

Fig. 5. Central wavelength (blue squares) and peak level (red triangles) of the clockwise output with 10-min interval.

The laser kept operating unidirectional, while the pulse feature was determined by the specific cavity dynamics. The generated pulses could be dechirped to 1.14 ps and 815 fs, which are longer when comparing to one of the most successful dissipative soliton lasers [2]. The strong pedestal of the compressed pulse we believe is due to the triangle-top spectrum with a small 3-dB bandwidth and the steep edge. To further investigate the energy scalability of this laser, we consider inducing a concrete single passband spectral filter to suppress the Raman scattering and extending the laser into large normal dispersion regime searching for larger pulse energy and shorter pulse duration. At the meantime, enhance the cavity directional nonlinearity asymmetry for stable unidirectional operation.

## 4. Conclusion

In this paper, we have experimentally demonstrated self-started unidirectional DS operation and NLP operation in the all-normal-dispersion fiber ring lasers mode locked with the NPR technique without an isolator. The cavity directional nonlinearity asymmetry caused the CW direction mode locked first. After mode locking, the laser operated unidirectional due to the NPR mode locking technique. The laser shown good operation stability and promising energy scalability. The results suggest that one could define the laser operation direction without an isolator by adopting the NPR mode locking and inducing large cavity directional asymmetry through cavity design. Large cavity nonlinearity discrimination will become more and more feasible as the laser pulse power keeps increasing, which will ultimately result in mode locking.

### Acknowledgements

This work was supported in part by the National Natural Science Foundation of China under Grant 61177045, 11274144, 61275109 and 61405079, in part by the Priority Academic Program Development of Jiangsu higher education institutions (PAPD), in part by the the Jiangsu Province Science Foundation (BK20140231).